\documentclass{amsart}

\usepackage{amsmath}    % Enables the align enviroment
\usepackage{amssymb}    % Math symbols (e.g. \mathbb{})
\usepackage{dsfont} 	% For \mathds{1} blackboard bold 1
\usepackage{bm}         % For roman (upright) bold latin letters
\usepackage{mathtools}  % Better math
\usepackage[hypertexnames=false]{hyperref} % For urls and hyperlinks
\usepackage[scientific-notation=true]{siunitx}
\usepackage{enumerate}
\usepackage{centernot}
\usepackage{pdflscape}
\usepackage{paralist}
\usepackage{stackrel}
\usepackage[english]{babel}
\usepackage[table]{xcolor}
\usepackage{setspace}

\usepackage[numbers]{natbib}
\usepackage{bibunits}

\usepackage{algorithm}
\usepackage{algpseudocode}

\usepackage{listings}
\usepackage{color}

\newtheorem*{heuristic*}{Heuristic Definition}
\theoremstyle{definition}
\makeatletter
\newcommand{\addresseshere}{%
  \enddoc@text\let\enddoc@text\relax
}

\definecolor{lightgrey}{rgb}{0.9,0.9,0.9}
\definecolor{darkgreen}{rgb}{0,0.6,0}
\begin{document}

%%%%%%%%%%%%%%%% FRONT %%%%%%%%%%%%%%%%%%%%%%%%%%%%%%%%%%%%%%%%%%%%%%
%--------------- Title & Abstract -----------------------------------
\title[Short version of the Amsterdam IADL Questionnaire]
{Detecting functional decline \\ from normal ageing to dementia: \\
Development and validation of a short version of \\ the Amsterdam IADL Questionnaire$^\copyright$}

\author[R.J.\ Jutten et al.]{Roos J.\ Jutten*}
\thanks{*Corresponding author.}
\address[Roos J.\ Jutten]{
Alzheimer Center and Dept.\ of Neurology \\
Amsterdam Neuroscience \\
VU University Medical Center \\
Amsterdam \\
The Netherlands}
\email{r.jutten@vumc.nl}

\author[]{Carel F.W.\ Peeters}
\address[Carel F.W.\ Peeters]{
Dept.\ of Epidemiology \& Biostatistics \\
Amsterdam Public Health research institute \\
VU University medical center \\
Amsterdam\\
The Netherlands}
\email{cf.peeters@vumc.nl}

\author[]{Sophie M.J.\ Leijdesdorff}
\address[Sophie M.J.\ Leijdesdorff]{
Alzheimer Center Rotterdam \\
Erasmus Medical Center \\
Rotterdam \\
The Netherlands}
\email{s.leijdesdorff@erasmusmc.nl}

\author[]{Pieter Jelle Visser}
\address[Pieter Jelle Visser]{
Alzheimer Center and Dept.\ of Neurology \\
Amsterdam Neuroscience \\
VU University Medical Center \\
Amsterdam \\
The Netherlands; \and
Alzheimer Center \\
School for Mental Health and Neuroscience \\
University Medical Centre Maastricht \\
Maastricht \\
The Netherlands}
\email{pj.visser@maastrichtuniversity.nl}

\author[]{Andrea B.\ Maier}
\address[Andrea B.\ Maier]{
MOVE Research Institute Amsterdam \\
Department of Human Movement Sciences \\
VU University Amsterdam \\
Amsterdam \\
The Netherlands; \and
Dept.\ of Medicine and Aged Care \\
Royal Melbourne Hospital \\
University of Melbourne \\
Melbourne \\
Australia}
\email{a.b.maier@vu.nl}

\author[]{Caroline B.\ Terwee}
\address[Caroline B.\ Terwee]{
Dept.\ of Epidemiology \& Biostatistics \\
Amsterdam Public Health research institute \\
VU University medical center \\
Amsterdam \\
The Netherlands; \and
The EMGO Institute for Health and Care Research \\
VU University Medical Center \\
Amsterdam \\
The Netherlands}
\email{cb.terwee@vumc.nl}

\author[]{Philip Scheltens}
\address[Philip Scheltens]{
Alzheimer Center and Dept.\ of Neurology \\
Amsterdam Neuroscience \\
VU University Medical Center \\
Amsterdam \\
The Netherlands}
\email{p.scheltens@vumc.nl}

\author[]{Sietske A.M.\ Sikkes}
\address[Sietske A.M.\ Sikkes]{
Alzheimer Center and Dept.\ of Neurology \\
Amsterdam Neuroscience \\
VU University Medical Center \\
Amsterdam\\
The Netherlands; \and
Dept.\ of Epidemiology \& Biostatistics \\
Amsterdam Public Health research institute \\
VU University medical center \\
Amsterdam \\
The Netherlands}
\email{s.sikkes@vumc.nl}

\begin{abstract}\label{abstract}

~\\
\noindent\emph{Introduction:}
Detecting functional decline from normal ageing to dementia is relevant for diagnostic and prognostic purposes.
Therefore, the Amsterdam IADL Questionnaire (A-IADL-Q) was developed: A 70-item proxy-based tool with good psychometric properties.
We aimed to design a short version whilst preserving its psychometric quality.

\noindent\emph{Methods:}
Study partners of subjects ($n=1355$), ranging from cognitively normal to dementia subjects, completed the original A-IADL-Q.
We selected the short version items using a stepwise procedure combining missing data, Item Response Theory and input from respondents and experts.
We investigated internal consistency of the short version as well as concordance with the original version.
To assess its construct validity, we additionally investigated concordance between the short version and the Mini-Mental State Examination (MMSE) and Disability Assessment for Dementia (DAD).
Lastly, we investigated differences in IADL scores between diagnostic groups across the dementia spectrum.

\noindent\emph{Results:}
We selected 30 items covering the entire spectrum of IADL functioning.
Internal consistency (.98) and concordance with the original version (.97) were very high.
Concordance with the MMSE (.72) and DAD (.87) scores was high.
IADL impairment scores increased across the spectrum from normal cognition to dementia.

\noindent\emph{Discussion:}
The A-IADL-Q Short Version (A-IADL-Q-SV) consists of 30 items.
The A-IADL-Q-SV has maintained the psychometric quality of the original A-IADL-Q.
As such, it is a concise measure of functional decline.

\begin{sloppypar}

\bigskip
\noindent \emph{Abbreviations}:
AD = Alzheimer's disease;
A-IADL-Q = Amsterdam IADL Questionnaire;
CFI = Comparative fit index;
DAD = Disability Assessment for Dementia;
IADL = Instrumental activities of daily living;
IIC = Item information curve;
IRT = Item response theory;
GRM = Graded response model;
LRT = Likelihood ratio test;
MCI = Mild cognitive impairment;
MML = Marginal maximum likelihood;
MMSE = Mini-Mental State Examination;
NC = Normal cognition;
RMSEA = Root mean square error of approximation;
SCD = Subjective cognitive decline;
SD = Standard deviation;
VAS = Visual analogue scale;
VUmc = VU University Medical Center
\end{sloppypar}
\end{abstract}

\maketitle

%%%%%%%%%%%%%%%% MAIN %%%%%%%%%%%%%%%%%%%%%%%%%%%%%%%%%%%%%%%%%%%%%%%
%--------------- Introduction ---------------------------------------
\section{Background}\label{Intro}
Dementia is a syndrome characterized by progressive cognitive decline and significant interference in daily function [1].
The first observable problems in daily life often concern the Instrumental Activities of Daily Living (IADL).
IADL can be defined as `complex activities for which multiple cognitive processes are necessary', such as cooking, managing finances and driving [2].
Detecting functional decline along the continuum from normal ageing to dementia is highly relevant for a number of reasons.
First of all, subtle IADL problems may already be present in subjects with Mild Cognitive Impairment (MCI) and predict progression to dementia [3-5].
This suggests that assessment of IADL can be used to select MCI subjects at an increased risk for dementia [6].
Once a diagnosis has been established, measuring IADL performance remains essential for the monitoring of clinical progression [7].
Lastly, IADL assessment plays a pivotal role in clinical trials, particularly in the evaluation of symptomatic treatment in dementia due to Alzheimer's disease (AD) [8-10].

IADL performance is often measured using proxy-based questionnaires [11].
Unfortunately, most of these questionnaires suffer from serious limitations.
They focus on everyday activities that are outdated and less relevant for patients in the early stages of dementia [12].
Furthermore, psychometric properties such as reliability, validity and responsiveness are often questionable or overlooked [13].
Recent studies have pointed out that improvements in IADL instruments are necessary, especially for detecting IADL problems in MCI and the early stages of dementia [14-17].

To overcome the above mentioned drawbacks of existing IADL scales, Sikkes et al. developed the Amsterdam IADL Questionnaire (A-IADL-Q).
The A-IADL-Q is a 70-item proxy-based tool and was developed with input from clinicians, patients and caregivers [18].
Previous studies have reported good psychometric properties with respect to reliability, validity, responsiveness and diagnostic accuracy in dementia [19-21].
One disadvantage of the A-IADL-Q is its length, resulting in an administration time of 20-25 minutes.
Additionally, respondents often report that some items are redundant or unclear.
To facilitate its administration and implementation on a wider scale, we aimed to design a short and more concise version of the A-IADL-Q.

The present paper describes the development and validation of a short version of the A-IADL-Q.
We aimed to select the most informative items, using a combined approach of quantitative and qualitative methods.
We expected that the short version would maintain the good psychometric quality of the original A-IADL-Q.
Additionally, we expected that IADL scores based on the short version would differ between diagnostic groups across the spectrum from normal cognition to dementia.

%--------------- Methods ----------------------------------
\section{Methods}\label{Methods}
\subsection{Subjects}\label{Subjects}
We selected $1355$ subjects with different levels of cognitive functioning, ranging from normal cognition to dementia.
Their study partner, mainly a spouse, relative or friend, completed the A-IADL-Q.
We included subjects from neurological memory clinics of the VU University Medical Center (VUmc) Alzheimer Center, Amsterdam, The Netherlands ($n=1117$), and the Alzheimer Center Rotterdam, The Netherlands ($n=32$) and from the geriatric memory clinic of the VUmc, Amsterdam, The Netherlands ($n=102$).
All these subjects underwent a dementia assessment, including clinical history, medical and neurological examination, screening laboratory tests, a neuropsychological test battery and brain imaging [22].
During this visit, study partners completed the A-IADL-Q on an iPad. Subjects' diagnoses were made in a multidisciplinary diagnostic meeting, containing at least a neurologist or geriatrician[3,22,23].

\begin{sloppypar}
We included cognitively normal subjects ($n=104$) from the Amsterdam site of the preclinAD cohort of the European Medical Information Framework for Alzheimer's disease (EMIF-AD) project.
Inclusion criteria for this cohort were: age $\geq 60$, Modified Telephone Interview for Cognitive Screening $> 22$; Geriatric Depression Scale $< 11$; Consortium to Establish a Registry for Alzheimer's Disease $10$ word list delayed recall $> 1.5$ SD of age adjusted normative data; and Clinical Dementia Rating score of $0$ with a score on the memory sub-domain of $0$ [24-27].
During the baseline visit, study partners completed the A-IADL-Q on an iPad.
\end{sloppypar}

Data were collected between October 2012 and August 2015.
All subjects gave written informed consent and all study partners gave oral informed consent.
The Medical Ethical Committee of the VU University Medical Center approved the study.

\subsection{The Amsterdam IADL Questionnaire}\label{IADLQ}
\begin{figure}[h]
\centering
  \includegraphics[width=\textwidth]{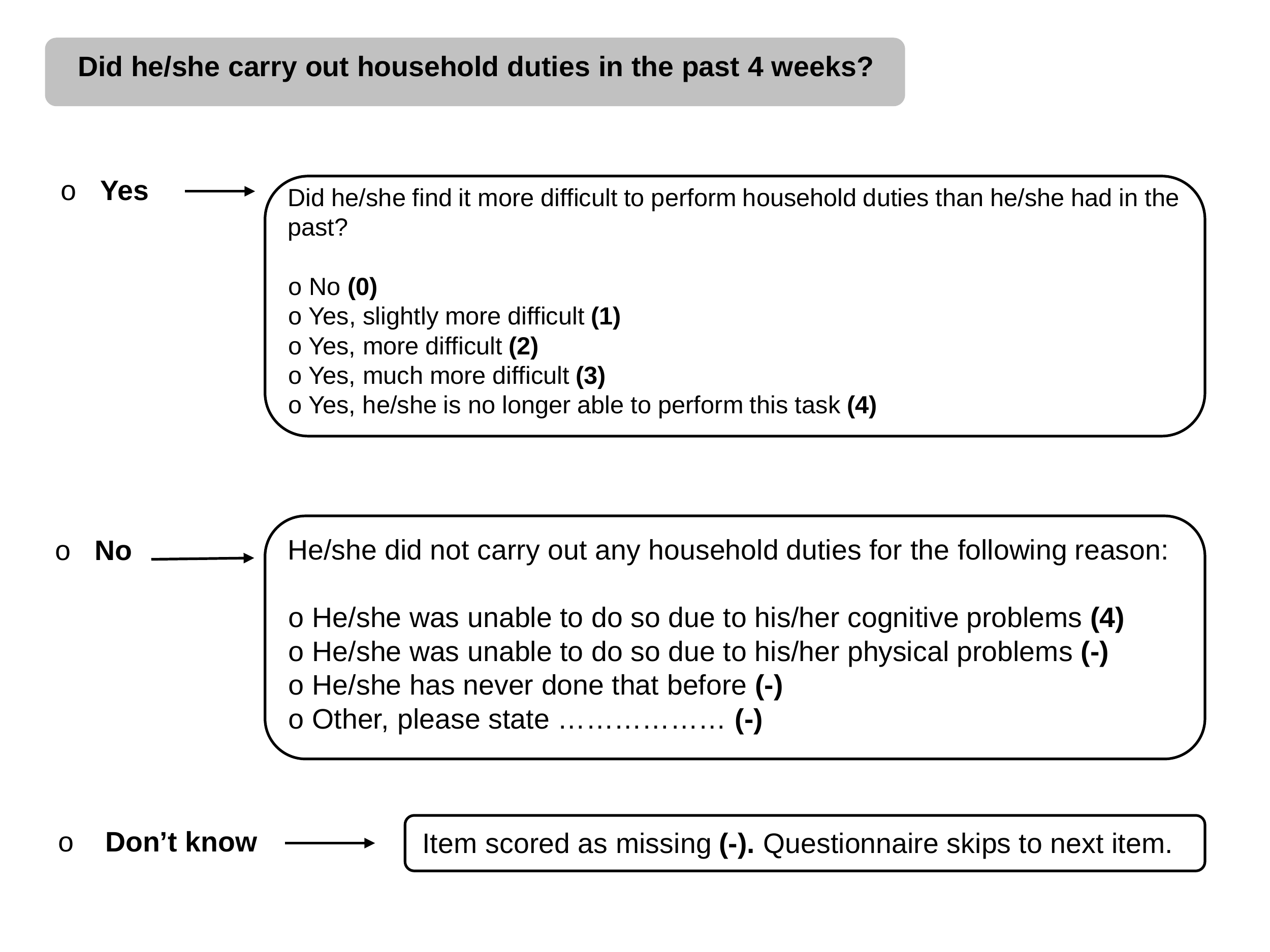}\\
  \caption{Example item of the A-IADL-Q, including response options and scoring.}
  \label{Fig:fig1EQ}
\end{figure}

The original A-IADL-Q is a proxy-based scale with $70$ items covering a broad range of cognitive IADL [18].
The items can be divided into eight subcategories: household, administration, work, computer use, leisure time, appliances, transport and other activities.
The A-IADL-Q is computerized and has an adaptive approach as the items are tailored to individual responses (see Figure \ref{Fig:fig1EQ}).
This results in a minimum of $47$ and a maximum of $70$ items for each respondent.
Prior to the start, it is emphasized that the questionnaire addresses day-to-day problems caused by cognitive problems, such as memory, attention, or planning problems.
Difficulty in performance is rated on a $5$-point Likert scale, ranging from `no difficulty in performing this task' to `no longer able to perform this task'.
Scoring is based on Item Response Theory (IRT): a paradigm linking responses to a test battery to an underlying construct (or latent trait) [28].
For the A-IADL-Q, the construct underlying the items can be termed `IADL performance'.
That is, the latent trait reflects IADL impairment with higher estimated trait levels indicating more impairment.

Linking the probabilities of category-specific item responses to latent trait levels is based on an IRT model [28].
For the A-IADL-Q, the Graded Response Model (GRM) is used: a polytomous IRT model appropriate for items with ordinal response categories [29].
In the GRM, each item is characterized by a discrimination parameter ($\alpha$) and 4 extremity parameters ($\beta$'s; the number of response categories minus 1).
The discrimination (or slope) parameter indicates how well an item discriminates between individuals with differing trait levels: higher discrimination parameters suggest higher ability to differentiate.
The extremity (or category threshold) parameters represent the trait levels that mark the transition between response categories (in terms of cumulative probabilities for endorsement) [29].
An important advantage of IRT is that one's level of the latent trait can be estimated from any set of items for which the parameters are known.
Therefore, IRT is able to handle missing data that may result from an adaptive approach.
IRT is often preferred over classical scoring methods for scale development and refinement: it advances the development of more efficient scales by supporting item-reduction whilst preserving measurement precision [30,31].

The following basic assumptions underlie the IRT framework: (1) Unidimensionality, implying that a single latent trait underlies the items; (2) Local independence, which implies independence of item responses conditional on the latent trait; and (3) Monotonicity, implying that the probability of endorsing (a category-specific response to) an item should increase as the trait level increases [32]. Previous work showed that the A-IADL-Q could be adequately described by a single latent factor and that the assumptions of local independence and monotonicity were met as well [19].
Since the current study contains a larger and more heterogenic sample, we have assessed these basic assumptions again.

\subsection{Procedures}\label{Procedures}
We divided the total sample into a training ($n=677$) and validation set ($n=678$), to use independent samples for the development and validation of the short version.
We randomly split the Alzheimer Center Rotterdam, the VUmc geriatric and the cognitively healthy cohorts.
We conducted an alternative split procedure for the VUmc Alzheimer Center cohort ($n=1117$), as a subsample ($n=206$) of this cohort was used for the validation of the original A-IADL-Q.
We therefore assigned this entire subsample to the current training set.
From the remaining subjects ($n=911$) we randomly assigned 35\% to our training set and 65\% to our validation set, to ensure that both sets had equal group sizes.

\subsubsection{Development procedure}\label{DevelProc}
Item selection was performed in the training set, using a stepwise procedure that combined missing data, IRT and content aspects.
As shown in Figure \ref{Fig:fig1EQ}, a response is scored as missing when (1) the particular task has not been performed due to other reasons rather than cognitive problems or (2) the study partner does not know whether the subject has performed that particular task in the past four weeks.
Items with higher percentages of missing responses give us a less direct view of cognitive IADL, and are thus less applicable for our goal.
We therefore eliminated items with more than 80\% overall missing data.
Items with more than 60\% missing data in all diagnostic groups were candidates for elimination.

\emph{IRT analyses}. We explored whether all items met the basic assumptions for IRT and eliminated items that did not meet these conditions.
In the subsequent refitting rounds, we used IRT to identify items that contributed little unique information to the model, as reflected in either low item information values (an index representing the precision with which the trait is measured) or overlapping Item Information Curves (IIC's; a mapping of the item information to the domain of the trait indicating how the information is distributed over the trait) with other items.
After each elimination round, the GRM was refitted and an overall fit-assessment was performed.
This resulted in new item parameters and IIC's that were used in the succeeding refitting round.

\emph{Content aspects}. Comprehensibility was investigated in two ways: (1) by inspecting the comments that respondents provided in the `comment box' after completing the A-IADL-Q; and (2) performing thinking-out-loud interviews in a subsample of respondents ($n=17$) whilst they were completing the A-IADL-Q.
Items that were often commented as unclear or redundant, in either the comment box or interview, were candidates for removal.
Furthermore, we investigated relevance and cultural applicability of all A-IADL-Q items with an online survey that we distributed among international experts.
Between February 2016 and May 2016, we distributed the survey through contacts of the authors (R.J.J. and S.A.M.S.) via Qualtrics (\url{www.qualtrics.com}).
All respondents ($n=33$) were clinicians or researchers representing $7$ countries, and had experience with the administration or cross-cultural validation of the A-IADL-Q.
They were asked to rate the necessity of each original A-IADL-Q item for inclusion in the short version on a visual analogue scale (VAS) ranging from $0$ (`not necessary at all') to $100$ (`very necessary').

\subsubsection{Validation procedure}\label{ValidProc}
To confirm the quality of the final short version, we investigated missing data patterns, experts' ratings, adherence to IRT assumptions, as well as the overall fit of the short version items in the validation set.
We subsequently investigated internal consistency of the short version and concordance between sum scores derived from the short and original version.
To assess construct validity [33], we investigated the relationship between the short version and measures of global cognition (Mini-Mental State Examination; MMSE [34]) and daily function (Disability Assessment for Dementia; DAD [35]), which were available for the VUmc Alzheimer Center cohort.
Based on previous results [19], we expected moderate-to-high concordance between the short version and MMSE and DAD scores.
To assess interpretability of the short version, we investigated differences in scores between six diagnostic groups that should represent different trait levels: (1) normal cognition (NC); (2) subjective cognitive decline (SCD); (3) mild cognitive impairment (MCI); (4) dementia due to AD (AD dementia); (5) dementia other than AD (non-AD dementia); and (6) another neurological or psychiatric disorder than dementia (Other).

\subsection{Statistical analyses}\label{StatAnal}
Statistical analyses were performed using \texttt{R} and \texttt{SPSS} version 20.0 [36,37].
Statistical significance (for multiplicity corrections) was set at $p <.05$.

\subsubsection{Development analyses}\label{DevelAnal}
Item selection was partly based on IRT modeling.
We used a GRM with a logit link function [29].
This model was fitted on the basis of approximate marginal maximum likelihood (MML) estimation [38].
The latent trait was assumed to follow a standard normal distribution.
We assessed unidimensionality by performing an eigenvalue decomposition on the matrix of robust (Spearman) correlations between the items.
A difference approximation to the second-order derivatives along the eigenvalue curve (scree plot) was calculated.
This acceleration-approximation indicates points of abrupt change along the eigenvalue curve [39].
The number of eigenvalues before the point with the most abrupt change (the point with the maximum acceleration value) represents the number of latent dimensions that dominate the information content. Local independence was assessed by inspecting residual correlation matrices.
We considered residual correlations above .25 as indicative of problematic item pairs.
We evaluated the monotonicity assumption using Mokken scale analysis [40].
Items that gave at least 1 significant violation of manifest monotonicity and had a crit value over 30 were considered to violate latent monotonicity [41].
We assessed basic model fit by comparing nested models: we employed a likelihood ratio test (LRT) to evaluate if the full GRM provided a better fit than a constrained GRM with equal slope parameters across items [28].

\subsubsection{Validation analyses}\label{ValidAnal}
We fitted a GRM on the final set of retained items.
Estimation and assumption evaluation for this model were performed as described above.
This model was also compared to a constrained GRM as a means of basic model fit assessment.
In addition, we evaluated global fitness of the final model with the comparative fit index (CFI) and root mean square error of approximation (RMSEA) [42].
Trait (or factor) scores were then based on empirical Bayes estimates: the mode of the posterior distribution of the trait given the retained items evaluated at the MML estimates.
We calculated internal consistency of the retained items using a robust version of McDonald's omega [43].
We examined concordance between sum-scores derived from the short and original versions, as well as between short version sum-scores and MMSE and DAD scores, using Kendall's $W$ [44].
To assess whether the short version scores differed between the diagnostic groups, we used a Kruskal-Wallis rank sum test on the trait scores followed by Dunn's pairwise test for multiple comparisons of mean rank sums (a nonparametric alternative to ANOVA followed by post-hoc tests) [45].
Multiple testing correction was based on the Bonferroni method.

%--------------- Results ----------------------------------
\section{Results}\label{Results}
\subsection{Sample and item characteristics}\label{Character}
The study sample consisted of subjects with NC ($n=104$), SCD ($n=219$), MCI ($n=138$), AD dementia ($n=413$), non-AD dementia ($n=235$) and $246$ subjects with other diagnoses.
Table \ref{Table:Character} shows subject characteristics for the total sample and for the training and validation set separately.
There were no age and gender differences between the two sets.
The MCI group was slightly larger in the training set, whilst the non-AD dementia group was slightly larger in the validation set.

Missing responses on item level in the training set ranged from 10.5\% (`preparing sandwich meals') to 92.8\% (`programming a video recorder').
Approximately half of the original version items (36/70) contained more than 50\% missing data.
Mean ratings from the 33 experts ranged from 23.9 (`programming a video recorder') to 86.9 (`paying when doing the shopping'), with an overall mean score of 62.3 (SD = 14.9).

\begin{table}[]
\centering
\caption{Subject characteristics.
Abbreviations: NC = normal cognition, SCD = subjective cognitive decline, MCI = mild cognitive impairment, AD = Alzheimer's disease.
$^\dag$ Tested using independent $t$-test.
$^\ddag$ Tested using Pearson's Chi-Square test.}
\label{Table:Character}
\resizebox{\textwidth}{!}{
\begin{tabular}{llllll}
\hline
                    &                 & Total sample& Training set & Validation set  & $p$-value \\
                    &                 & ($n=1355$)              &  ($n=677$) &  ($n=678$) &  \\ \hline
Age, M (SD)         &                 & 65.7 (9.7)            & 66.1 (10.1)           & 65.3 (9.2)             & 0.146$^\dag$  \\
Gender, female (\%) &                 & 602 (44.4\%)          & 301 (44.5\%)         & 301 (44.4\%)           & 0.981$^\ddag$  \\
Diagnosis           &                 &                       &                      &                        &         \\
                    & NC              & 104 (7.7\%)           & 52 (7.7\%)           & 52 (7.7\%)             &         \\
                    & SCD             & 219 (16.2\%)          & 116 (17.1\%)         & 103 (15.2\%)           &         \\
                    & MCI             & 138 (10.2\%)          & 84 (12.4\%)          & 54 (8.0\%)             &         \\
                    & AD dementia     & 413 (30.5\%)          & 209 (30.9\%)         & 204 (30.1\%)           &         \\
                    & non-AD dementia & 235 (17.3\%)          & 100 (14.8\%)         & 135 (19.9\%)           &         \\
                    & Other           & 246 (18.2\%)          & 116 (17.1\%)         & 130 (19.2\%)           &         \\ \hline
\end{tabular}
}
\end{table}

\subsection{Development of the short version}\label{Short}
\begin{figure}[h]
\centering
  \includegraphics[width=\textwidth]{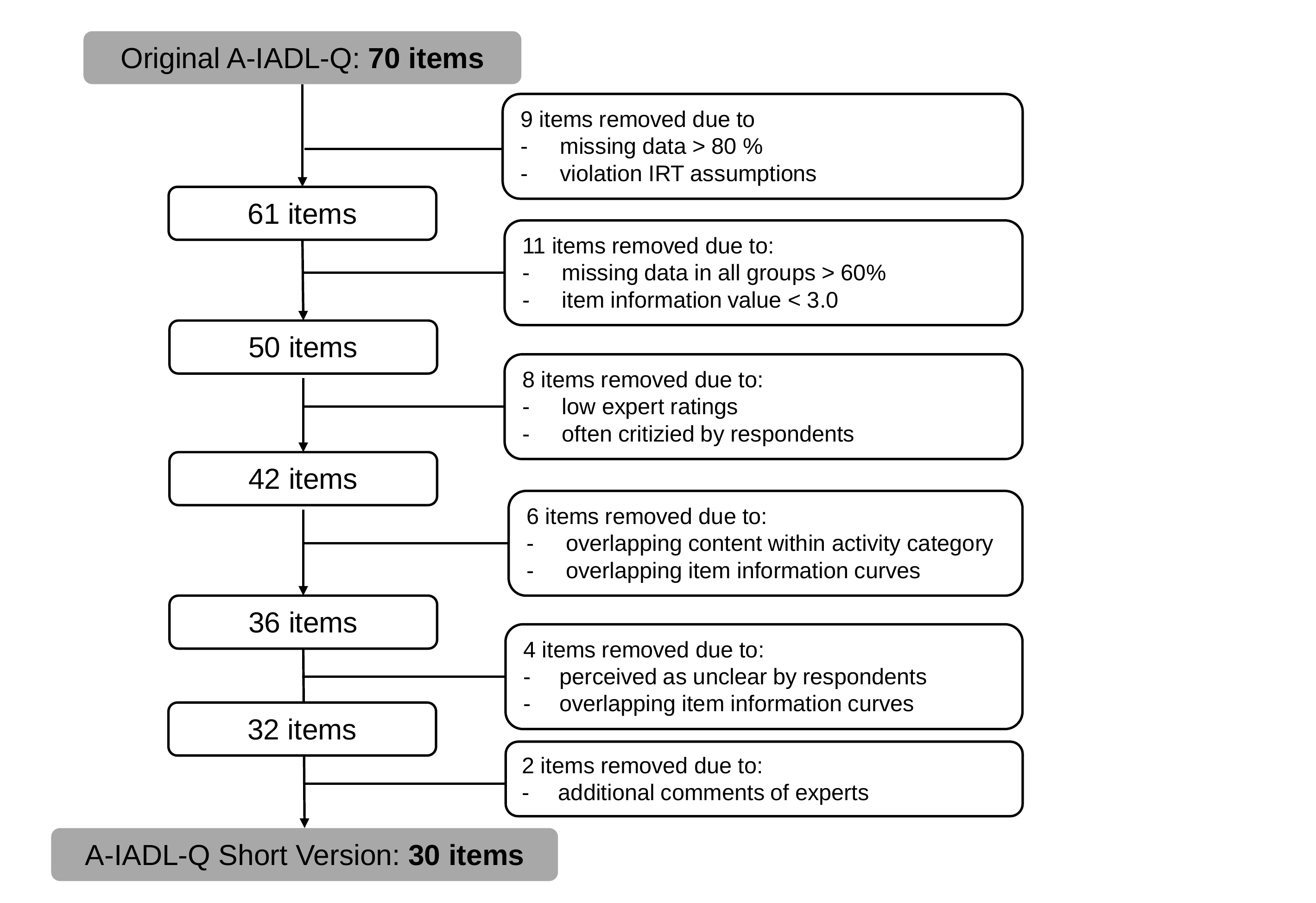}\\
  \caption{Flowchart of the item selection procedure that led to the short version of the Amsterdam IADL Questionnaire (A-IADL-Q). Abbreviation: IRT = Item Response Theory.}
  \label{Fig:fig2Flow}
\end{figure}

Figure \ref{Fig:fig2Flow} provides a flow-chart of the item selection procedure.
Our first step included the removal of $2$ items that violated the assumption of monotonicity, together with $7$ items that contained more than 80\% missing data.
After the second round, we removed $11$ items with missing responses above 60\% in all diagnostic groups and contributing little information to the model (item information $< 3.0$).
After the third round, we removed $8$ items that received low ratings of experts (mean rating $< 50$) and had often been commented on as either unclear or redundant by respondents.
We thereafter removed $6$ items with overlapping IIC's and overlapping content with other items within the same activity category (e.g. `cooking' versus `preparing hot meals').
Of these overlapping pairs, we removed the one containing higher missing data and lower content rating.
After the fourth round, we removed $4$ items that were often perceived as unclear and showed overlapping IIC's with more specific items (e.g., `looking for important things at home' versus `looking for his/her keys').
Lastly, we removed $2$ items due to disputable item characteristics and additional comments of experts.
Following this, we refitted the model with the remaining $30$ items and concluded that further shortening was unnecessary.
All $30$ retained items in the training set were deemed to contribute substantially unique information to the latent trait.
The full GRM model improved fit upon the constrained GRM model (LRT value = $98.01$, df = $29$, $p <.001$).

\subsection{Validation of the short version}\label{ValidShort}
\begin{figure}[t]
\centering
  \includegraphics[width=\textwidth]{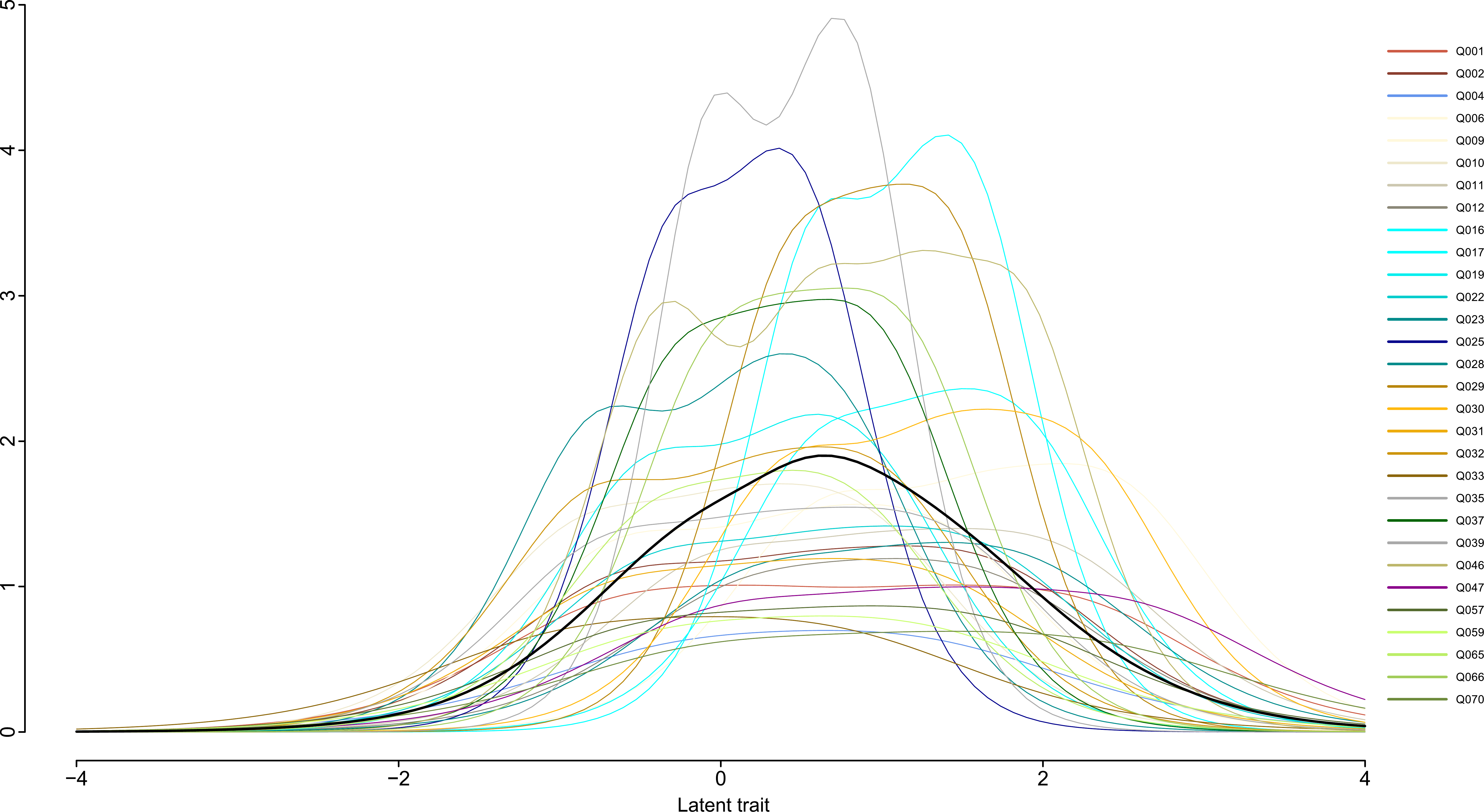}\\
  \caption{Item information curves of the 30 Amsterdam IADL items that resulted in the short version.
  The bold black line represents the total test information curve.
  The latent trait ranges from $-4$ (good IADL functioning) to $+4$ (poor IADL functioning).}
  \label{Fig:fig3Info}
\end{figure}

The final selection of items can be found in Column 1 of Table \ref{Table:GRM}.
This selection adheres to all assumptions underlying the IRT framework.
The maximum acceleration factor on the consecutive eigenvalues of the robust correlation matrix occurs at the second eigenvalue (with a value of $1.26$), implying that the first eigenvalue (with a value of $17.09$) dominates the information content.
Hence, a single latent dimension is sufficient.
Moreover, no item pair sorted a residual correlation above $.25$ and no item displayed significant violations of manifest monotonicity.
Table \ref{Table:GRM} also presents information on missing percentages and the estimated GRM parameters based on the validation set.
The last column shows the experts' ratings.
As can be seen, all retained items contain less than 60\% missing data and most items (26/30) had less than 50\% missing data in the validation set.
The extremity parameters were spread along the latent trait continuum (ranging from $-4$ to $+4$), which is also illustrated by the IIC's presented in Figure \ref{Fig:fig3Info}.
For most items, item information values were above $3$ (on a total information of $163.68$).
Lastly, all short version items received medium to high ratings from experts.

\begin{table}[]
\centering
\caption{Final selection of the short version items, including their missing data percentages, GRM parameters, item information values and content ratings by experts.
Abbreviations: GRM = Graded Response Model, $\alpha$ = discrimination parameter, $\beta$'s = extremity parameters.
NOTE: Percentage missing, parameter estimates and information characteristics are based on the validation set.
Expert ratings were made per item on a visual analogue scale ranging from 0 to 100.}
\label{Table:GRM}
\resizebox{\textwidth}{!}{
\begin{tabular}{llrrrrrrrr}
\hline
Item &                                              & \% Missing &       &        & Item  &       &       & Item  & Expert  \\
     &                                              &            &       &        & param.                &       &       & information  &  rating       \\
     &                                              &            &       &        &                 &       &       &                  &               \\ \cline{4-8}
     &                                              &            & $\alpha$     & $\beta_1$     & $\beta_2$              & $\beta_3$    & $\beta_4$   &                  &               \\ \hline
1    & Carrying out household duties	    		& 12.1       & 1.830  & -0.653 & 0.259	          & 1.309 & 2.209 & 4.73             & 76 \\
2    & Doing the shopping	            		& 14.3       & 2.027  & -0.597 & 0.481	          & 1.206 & 1.732 & 4.93             & 79 \\
4    & Buying the correct articles	    		& 34.8       & 1.505  & -0.209 & 0.795	          & 1.249 & 1.282 & 2.57             & 73 \\
6    & Cooking	                            		& 33.8       & 2.236  & -0.613 & 0.426	          & 0.990 & 1.388 & 5.25             & 76 \\
9    & Preparing sandwich meals	    			& 10.0       & 2.454  & 0.643  & 1.602	          & 2.266 & 2.524 & 5.82             & 60 \\
10   & Making minor repairs to the house   		& 53.1       & 2.330  & -0.867 & -0.013	          & 0.510 & 0.897 & 5.28             & 60 \\
11   & Operating domestic appliances	    		&  8.3       & 2.120  & -0.074 & 0.865	          & 1.544 & 2.142 & 5.17             & 63 \\
12   & Operating the microwave oven	    		& 30.4       & 1.938  & 0.134  & 0.880	          & 1.351 & 1.711 & 3.83             & 58 \\
16   & Operating the coffee maker	    		& 11.4       & 2.751  & 0.598  & 1.302	          & 1.731 & 1.913 & 5.82             & 63 \\
17   & Operating the washing machine	    		& 42.0       & 3.662  & 0.584  & 1.254	          & 1.533 & 1.610 & 7.71             & 58 \\
19   & Paying bills	                    		& 34.4       & 2.659  & -0.467 & 0.417	          & 0.717 & 0.904 & 5.55             & 83 \\
22   & Using a mobile phone	            		& 17.0       & 2.126  & -0.416 & 0.531	          & 1.148 & 1.658 & 5.00             & 76 \\
23   & Managing the household budget	    		& 45.0       & 2.884  & -0.773 & 0.138	          & 0.448 & 0.738 & 6.56             & 79 \\
25   & Using electronic banking	    			& 46.9       & 3.632  & -0.320 & 0.285	          & 0.500 & 0.560 & 7.12             & 66 \\
28   & Using a pin code	            			& 11.8       & 2.030  & 0.190  & 1.082	          & 1.533 & 1.984 & 4.34             & 77 \\
29   & Obtaining money from a cash machine 		& 32.7       & 3.486  & 0.397  & 0.892	          & 1.316 & 1.484 & 7.59             & 69 \\
30   & Paying using cash	            		& 15.6       & 2.665  & 0.415  & 1.288	          & 1.712 & 2.257 & 6.59             & 72 \\
31   & Making appointments	            		& 17.4       & 1.945  & -0.660 & 0.270	          & 1.363 & 1.945 & 4.33             & 75 \\
32   & Filling in forms	            			& 24.8       & 2.516  & -0.792 & 0.173	          & 0.754 & 1.307 & 5.95             & 66 \\
33   & Working	                            		& 47.1       & 1.579  & -0.965 & -0.237	          & 0.379 & 0.742 & 2.95             & 70 \\
35   & Using a computer	            			& 22.6       & 2.229  & -0.718 & 0.193	          & 0.846 & 1.418 & 5.47             & 68 \\
37   & Emailing	                    			& 44.0       & 3.100  & -0.297 & 0.293	          & 0.789 & 1.024 & 7.01             & 54 \\
39   & Printing documents	            		& 56.6       & 4.080  & -0.059 & 0.653	          & 0.814 & 0.857 & 8.3              & 70 \\
46   & Operating devices	            		& 16.4       & 3.374  & -0.367 & 0.584	          & 1.243 & 1.869 & 10.3             & 72 \\
47   & Operating the remote control	    		&  3.1       & 1.786  & 0.017  & 1.073	          & 1.734 & 2.599 & 4.32             & 80 \\
57   & Playing card and board games	    		& 50.9       & 1.657  & -0.462 & 0.556	          & 1.194 & 1.662 & 3.49             & 62 \\
59   & Driving a car	                    		& 25.7       & 1.592  & -0.351 & 0.569	          & 1.015 & 1.300 & 2.93             & 76 \\
65   & Using a sat-nav system	            		& 51.6       & 2.421  & -0.394 & 0.384	          & 0.777 & 0.823 & 4.62             & 61 \\
66   & Using public transport	            		& 47.9       & 3.123  & -0.071 & 0.420	          & 0.878 & 1.200 & 7.01             & 83 \\
70   & Being responsible for his/her own medication     & 27.7       & 1.478  & -0.075 & 1.005	          & 1.685 & 2.361 & 3.16             & 82 \\
\hline
\end{tabular}
}
\end{table}

The full GRM model provided better fit on the validation data than the constrained GRM model (LRT value = 6644.47, df = 29, $p <.001$).
The overall fit of the final model was considered good: CFI = .994, RMSEA = .032.
Internal consistency of the short version was very high (robust McDonald's omega = .98).
Concordance between the item sum-scores of the short version and the original version also was very high (Kendall's $W$ = .97).
Concordance with the MMSE (Kendall's $W$ = .72) and DAD (Kendall's $W$ = .87) were high.

\begin{figure}[h]
\centering
  \includegraphics[width=\textwidth]{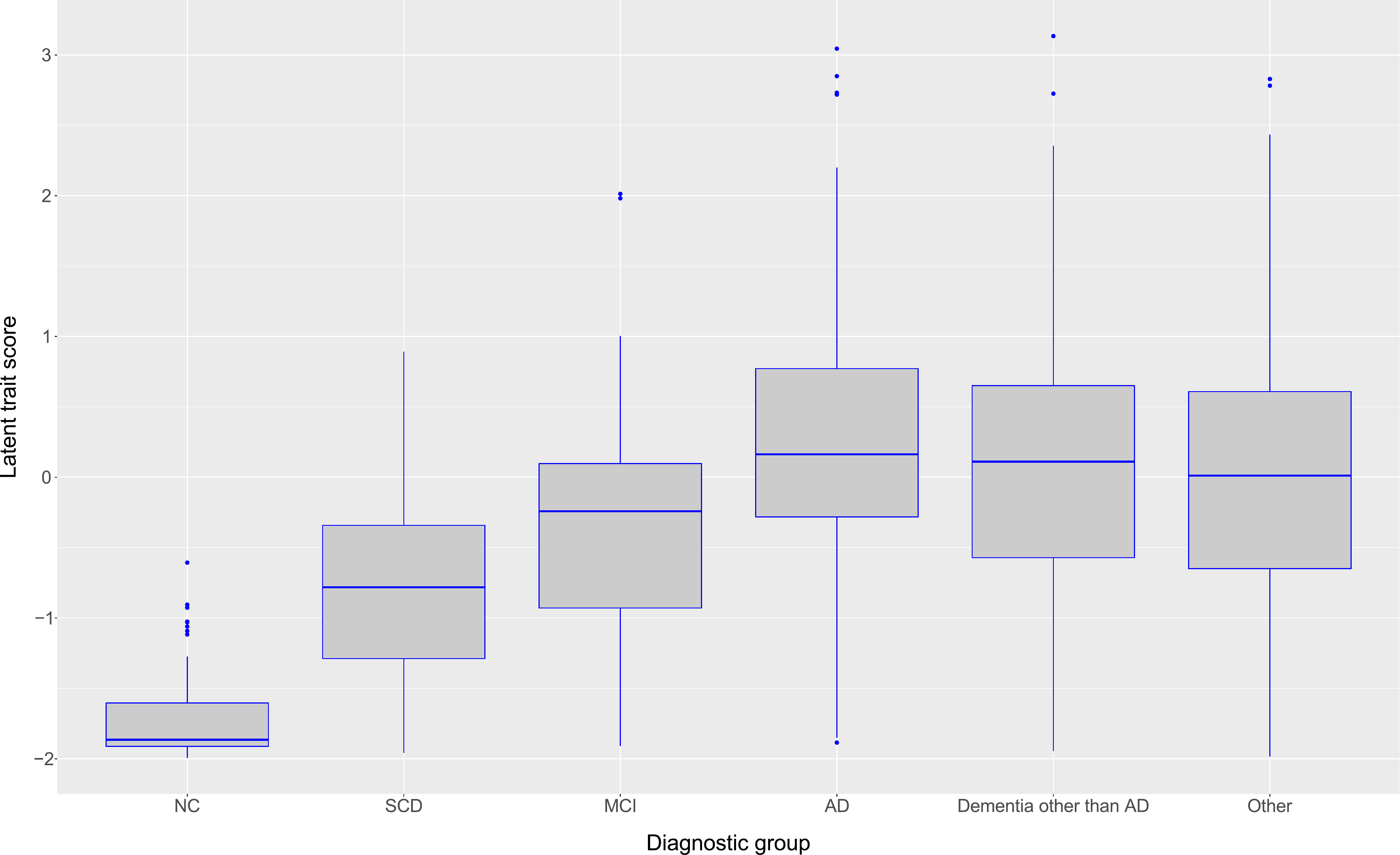}\\
  \caption{Short version latent trait scores for each diagnostic group.
  Latent trait scores reflect IADL functioning, with higher scores indicating poorer IADL functioning.
  Post-hoc analyses gave the following significant pairwise differences: 1) NC vs. all other groups; 2) SCD vs. AD dementia, non-AD dementia and Other group; and 3) MCI vs. AD dementia.
  Abbreviations: NC = normal cognition, SCD = subjective cognitive decline, MCI = mild cognitive impairment, AD = Alzheimer's disease.}
  \label{Fig:fig4Scores}
\end{figure}

Table \ref{Table:Clinpars} presents the clinical characteristics of the different diagnostic groups within the validation set.
Figure \ref{Fig:fig4Scores} represents the trait score distributions for each diagnostic group.
It can be seen that this score seems to increase from normal cognition to dementia.
The variances of the trait scores were not equal between diagnostic groups.
Hence, a nonparametric test was employed to assess diagnostic-group-differences between latent trait scores as derived from the final GRM model.
The Kruskal-Wallis rank sum test indicated that the mean trait score ranks of the diagnosis groups indeed differed ($\chi^{2}$ = 187.01, df = 5, $p <.001$).
Pairwise comparisons (Dunn's test) with Bonferroni correction indicated the following pairwise differences: (1) NC versus all other groups (all corrected $p$-values $< .001$); (2) SCD versus AD dementia, non-AD dementia and Other group (all corrected $p$-values $<.001$); and (3) MCI versus AD dementia (corrected $p$-value = .002).

\begin{table}[]
\centering
\caption{Clinical characteristics of different diagnostic groups in the validation set.
Abbreviations: NC = normal cognition, SCD = subjective cognitive decline, MCI = mild cognitive impairment,
AD = Alzheimer's disease, MMSE = Mini-Mental State Examination (higher scores reflect better cognitive functioning),
DAD = Disability Assessment for Dementia (lower scores reflect more dysfunction in activities of daily living),
N.A.= Not available for this cohort.}
\label{Table:Clinpars}
\resizebox{\textwidth}{!}{
\begin{tabular}{lllllll}
\hline
		& NC		& SCD		& MCI		& AD dementia		& Non-AD dementia		& Other  \\
		& $(n = 52)$	& $(n = 103)$	& $(n = 54)$	& $(n = 204)$		& $(n = 135)$			& $(n = 130)$  \\\hline
Female (\%)	& 30 (57.7\%)	& 43 (41.7\%)	& 18 (33.3\%)	& 111 (54.4\%)		& 57 (42.2\%)			& 42 (32.3\%) \\
Age (SD)	& 70.5 (7.6)	& 62.8 (10.2)	& 69.8 (9.3)	& 66.9 (8.9)		& 66.5 (8.5)			& 60.8 (9.9)  \\
MMSE score (SD)	& N.A.		& 27.3 (2.2)	& 26.6 (2.1)	& 20.1 (4.8)		& 23.7 (4.4)			& 25.3 (4.3)  \\
DAD score (SD)	& N.A.		& 91.0 (12.6)	& 86.5 (14.3)	& 78.2 (20.6)		& 75.7 (25)			& 76.6 (23.7) \\
\hline
\end{tabular}
}
\end{table}

%--------------- Discussion -------------------------------
\section{Discussion}\label{Discussion}
We designed a short version of the A-IADL-Q (A-IADL-Q-SV) containing 30 items.
We thereby reduced administration time by approximately 10 minutes.
We showed that, although significantly shorter, the A-IADL-Q-SV has maintained the psychometric quality of the original version.
We demonstrated adequate measurement precision along the entire spectrum of IADL functioning.
Short version scores were in high concordance with the MMSE and DAD, which supports the construct validity of the A-IADLQ-SV.
We also found that the A-IADL-Q-SV could differentiate between various diagnostic groups with respect to IADL impairment.

The current study expands on previous work on the A-IADL-Q, which already demonstrated good psychometric quality of the scale [18-21].
The A-IADL-Q-SV contains only the most informative items, and thereby possible `noise' caused by less informative or ambiguous items has been reduced.
Because of its reduced length, the A-IADL-Q-SV may be perceived as a more user-friendly measure.
The use of shorter tests is also encouraged from a psychometric point of view: a short form containing items of the same quality as the original form may yield less measurement error and thus be more reliable [28].
Longer tests are more likely to suffer from acquiescence bias and missing responses.
Using the A-IADL-Q-SV may overcome these test-length related drawbacks.

Our findings suggest that the A-IADL-Q-SV can already detect IADL problems in subjects with SCD and MCI, which is in line with previous studies that report subtle functional impairment in these groups [46-48].
We found that IADL scores differed between subjects with NC and SCD.
This is of particular interest since both groups are characterized by the absence of objective cognitive impairment, although SCD subjects may be at higher risk of developing dementia [49].
The A-IADL-Q-SV might thus be able to detect subtle functional decline that appears in preclinical stages of dementia, suggesting that it could be a promising measure for clinical trials in these earliest stages [10,50].

Strengths of this study include our large and heterogenic sample with subjects covering a broad range of the IADL spectrum along the continuum from normal ageing to dementia.
Another strength is the use of a validation set to replicate findings derived from the training set.
After splitting the total sample, the training and validation set both contained more than 500 subjects, a number that is recommended for estimating accurate parameters based on the GRM [51].
Finally, combining statistical methods with input from respondents and experts is an important strength of this study, as it preserved both the psychometric quality and clinical relevance of the A-IADL-Q-SV.

There are some limitations that should be considered.
Among them are our relatively small NC group, due to the fact that most subjects were recruited via memory clinics.
Secondly, previous studies have shown that proxy-based IADL measures may be confounded by respondent characteristics such as caregiver burden and depression [52].
We did not take these characteristics into account in the current study.
However, Sikkes et al. showed low correlations between the original A-IADL-Q, caregiver burden and depression, indicating limited confounding by these variables [19].

Further research is needed to examine whether the A-IADLQ-SV is sensitive to changes over time \emph{within} subjects.
We will investigate the A-IADL-Q-SV longitudinally in subjects with MCI and early dementia, in order to determine whether it could be an effective measure for monitoring disease progression and evaluating disease-modifying therapies.
Since the research field is shifting towards preclinical stages of dementia, it is also relevant to further investigate the A-IADL-SV in subjects with SCD and the relation between IADL scores and dementia biomarkers in this group.

To conclude, we developed a short version of the A-IADL-Q, which is a concise instrument to efficiently measure functional decline in the early stages of dementia.
The A-IADL-Q-SV has retained the good qualities of the original A-IADL-Q; hence, we expect the short version to be a promising outcome measure for daily function in dementia research as well as in clinical practice.

%%%%%%%%%%%%%%%% BACK %%%%%%%%%%%%%%%%%%%%%%%%%%%%%%%%%%%%%%%%%%%%%%%
%--------------- Thanks ---------------------------------------------
\section*{Acknowledgements}
The authors would like to thank all respondents and experts for their willingness to participate in this study.
We also would like to thank Naomi Koster, Saskia de Vries, Judith Meurs, Iris Dalhuizen and Tarik Binnenkade for their help with the data collection.
The development of the Amsterdam IADL Questionnaire is supported by grants from Stichting VUmc Fonds and Innovatiefonds Zorgverzekeraars.
The current study is supported by a grant from Memorabel (grant no. 733050205), which is the research programme of the Dutch Deltaplan for Dementia.
This work has received support from the EU/EFPIA Innovative Medicines Initiative Joint Undertaking (grant no. 115372).
Part of this paper has been presented at the 2016 AAIC conference.

The Amsterdam IADL Questionnaire$^{\copyright}$ is free for use in all public health and not-for-profit agencies and can be obtained via \url{https://www.alzheimercentrum.nl/professionals/amsterdam-iadl}.

%--------------- Disclosures ----------------------------------------
\section*{Conflict of interest}
R.J.J., C.F.W.P, S.M.J.L, P.J.V, A.B.M and C.B.T.\ report no relevant conflicts of interest.
P.S.\ has acquired grant support (for the institution; VUmc Alzheimer Center) from GE Healthcare, Danone Research, Piramal and MERCK.
In the past 2 years he has received consultancy/speaker fees (paid to the institution) from Lilly, GE Healthcare, Novartis, Sanofi, Nutricia, Probiodrug, Biogen, Roche, Avraham and EIP Pharma.
S.A.M.S.\ is supported by grants from JPND and Zon-MW, and has provided consultancy services in the past two years for Nutricia and Takeda.
All funds were paid to her institution.

%--------------- References -----------------------------------------

%--------------- Addresses ------------------------------------------
%\newpage
\vspace{.5cm}
\addresseshere

\end{document}